\documentclass{aastex61}
\usepackage{amsmath}

\received{}
\revised{}
\accepted{}
\submitjournal{ApJ}

\begin{document}

\title{A Circumbinary Disk Model for the Rapid Orbital Shrinkage in Black Hole Low-Mass X-ray Binaries}

\correspondingauthor{Xiang-Dong Li}
\email{lixd@nju.edu.cn}

\author{Xiao-Tian Xu}
\affil{Department of Astronomy, Nanjing University, Nanjing 210046, China}
\author{Xiang-Dong Li}
\affil{Department of Astronomy, Nanjing University, Nanjing 210046, China}
\affil{
Key Laboratory of Modern Astronomy and Astrophysics (Nanjing University), Ministry of Education, Nanjing 210046, China
}

\begin{abstract}
Several black hole low-mass X-ray binaries (BHLMXBs) show very fast
orbital shrinkage,  which is difficult to understand in the standard
picture of the LMXB evolution. Based on the possible detection of a
circumbinary (CB) disk in A0620-00 and XTE J1118+480, we investigate the
influence of the  interaction between a CB disk and the inner binary and
calculate the evolution of the binary using the Modules for Experiments in
Stellar Astrophysics. We consider two cases for the CB disk formation in
which it is fed by mass loss during single outburst or successive
outbursts in the LMXB. We show that when taking reasonable values of the
initial mass and the dissipating time of the disk, it is possible to
explain the fast orbital shrinkage in the BHLMXBs without invoking high
mass transfer rate.
\end{abstract}


\section{Introduction} \label{sec:intro}
Black hole low-mass X-ray binaries (BHLMXBs), which are a subclass of X-ray
binaries,  consist of a stellar-mass BH accretor and a low-mass  ($\lesssim
1\,M_{\sun}$) donor. There are 19 Galactic BHLMXBs dynamically confirmed
\citep{Rem2006, McC2006, Cas2014}. Most of  the donors are dwarf stars with
spectral types ranging from A2V to M1V. The masses of the BHs range from
$\sim 2.7\,M_\sun$ to $\gtrsim 15\,M_\sun$, and the orbital periods ($P_{\rm
orb}$) are usually $\lesssim 1$ day. All BHLMXBs are transients, alternating
between long quiescence with X-ray luminosity $L_{\rm X}
\sim10^{30.5}-10^{33.5}$ erg s$^{-1}$ and short outburst with $L_{\rm X}\sim
10^{37}-10^{39}$ erg s$^{-1}$. These different states are thought to be
related to different accretion processes taking place in BHLMXBs. The
quiescent and outburst states, which are characterized by low/hard and
high/soft X-ray spectra, can be explained by an advection-dominated
accretion flow \citep{Nar1995,Abr1995} and a optically thick, geometrically
thin disk \citep{Sha1973}, respectively; the transition between these states
is usually thought to be driven by the change in the accretion rate
\citep{Esi1997}.

In the standard model of the formation of BHLMXBs
\citep{vdH1983,Bha1991,Tau2006}, the progenitor  of a BHLMXB is a binary
with an extreme mass ratio in a relatively wide orbital \citep[see][for a
review]{Li2015}. After its main-sequence evolution, the primary star expands
rapidly and overflows its Roche lobe. Due to the high mass ratio the mass
transfer is dynamically unstable and the transferred material engulfs the
secondary star. Then, the system enters the common envelope (CE) phase
\citep{Pac1976}. The CE evolution yields a close binary system or merger of
the two stars, depending on the dissipation process in the stellar envelope
of the primary star. If the binary can survive the CE phase, the secondary
still remains in the main-sequence and the primary star evolves into a naked
core, which finally experiences core-collapse with the formation of a BH.  A
natal kick may be imparted to the BH during the collapse. If the binary can
survive, the subsequent evolution is driven by either nuclear evolution of
the secondary star or magnetic braking (MB) and gravitational radiation
(GR). Once the orbit is narrow enough for the onset of mass transfer, the
binary behaves as a LMXB.

Orbital evolution can provide useful probe of the mass transfer processes in
BHLMXBs. However,   it is challenging to measure the derivative
($\dot{P}_{\rm orb}$) of the orbital period in BHLMXBs. Recently,
\citet{Gon2012,Gon2014,Gon2017} reported the $\dot{P}_{\rm orb}$
measurements of three BHLMXBs (A0620-00, XTE J1118+480 and Nova Muscae 1991)
by using the Doppler shift of the companion's spectral lines, and found that
these BHLMXBs are experiencing extremely rapid orbital shrinkage (see Table
\ref{tab:data} for a summary of the observed parameters). The rates of
orbital decay exceed the expectation of the traditional theory invoking
angular momentum loss (AML) caused by MB and GR by around $1-2$ orders of
magnitude. If the orbital shrinkage is caused by mass loss, then the large
value of $|\dot{P}_{\rm orb}|$ requires that the donor be losing mass at a
very high rate so the binary would be bright in X-ray. However, the observed
X-ray luminosities of these BHLMXBs are all quite low: their
Eddington-scaled luminosities are about $10^{-3}-10^{-2}$
\citep{McC2006,Wu2010}.

This exotic behavior challenges our understanding about the evolution of
BHLMXBs. Several scenarios were suggested in the literature. (1) A massive
planet orbiting the binary may be responsible for the orbital shrinkage
\citep{Peu2014,Iar2015,Jai2017}. In this scenario, the orbit should
experience a series of shrinkage and expansion due to the motion of the
third-body. However, currently there is no direct observational indication
of the existence of a third-body in these BHLMXBs, and more observations are
needed to test this idea. (2) Some authors proposed that anomalous magnetic
activity of the low-mass donor may cause the rapid orbital decay (i.e., MB
with strong surface magnetic field of order 1 kG: \citealt{Jus2006} or the
Applegate mechanism: \citealt{Gon2014}). Evolutionary studies showed that
the effective temperatures of the donor stars are higher than observed in
the former model \citep{Jus2006}, and it remains to see whether the internal
energy budget of the low-mass donor star is sufficient to produce the
observed orbital period change with the Applegate mechanism in LMXBs
\citep{Pat2016}. (3) Based on the fact that there is a circumbinary (CB)
disk detected in both A0620-00 and XTE J1118+480 with mass
$\sim10^{22}-10^{24}$ g \citep{Mun2006}, \citet{Chen2015} investigated the
AM transfer between the CB disk and the binary caused by mass flow between
them. The mass flow itself can take away AM from the binary, causing the
orbit to shrink. However, to reach the observed shrinking rate, the expected
mass of the CB disk is six orders of magnitude higher than observed.

In this paper, we investigate an alternative CB disk model, considering
gravitational  resonant interaction between the disk and the binary.
Resonant interaction between a CB disk and the inner binary has been already
investigated in binary systems,  and it has been demonstrated that a CB disk
is able to effectively extract AM from the binary
\citep{Art1991,Lub1996,Lub2000,Der2013}. However, it is still open whether
this mechanism can account for the observation of BHLMXBs. The rest of this
paper is organized as follows. We describe the model for the CB disk-binary
interaction and the binary orbital evolution in section 2. Then we perform
numerical calculation of the binary evolution with different initial
parameters and evolutionary laws of the CB disk, and compare the results
with observations. In section 4, we summarize our work.

\section{Model}
In this work we assume that the observed orbital shrinkage is related to the
formation  and evolution of a CB disk, which originates from sporadic mass
loss process during outburst(s) of the BHLMXBs.  So, before investigating
the influence of the disk, we need to follow the long-term evolution of the
binary and set the initial condition at the onset of the CB disk formation.
We calculate the BHLMXB evolution with the Modules for Experiments in
Stellar Astrophysics (MESA; version number 9575)
\citep{Pax2011,Pax2013,Pax2015}. The physical considerations are described
as below.

The Roche-lobe $R_{\rm L,2}$ of the donor is evaluated with the formula proposed by \citet{Egg1983},
\begin{equation}
\frac{R_{\rm L,2}}{a}=\frac{0.49q^{2/3}}{0.6q^{2/3}+{\rm ln}(1+q^{1/3})},
\end{equation}
where $q=M_2/M_{\rm BH}$ is the mass ratio of the donor mass $M_2$ and the
BH mass  $M_{\rm BH}$, and $a$ is the orbital separation. We adopt the
Ritter scheme in MESA to calculate the mass transfer rate via Roche-lobe
overflow \citep{Rit1988}, 
\begin{equation}
-\dot{M}_2=\dot{M}_{2,0}{\rm exp}\left(-\frac{R_2-R_{\rm L,2}}{H}\right),
\label{mdot}
\end{equation}
where $R_2$ is the radius of the donor, $H$ is scale height of the
atmosphere evaluated  at the surface of the donor star, and
\begin{equation}
\dot{M}_{2,0}=\frac{1}{e^{1/2}}\rho c_{\rm th}Q,
\end{equation}
where $\rho$ and $c_{\rm th}$ are  the mass density and the sound speed at
the surface of the star respectively, and $Q$ is the cross section of the
mass flow via the L1 point. We assume that the accretion rate onto BH is
limited by the Eddington accretion rate $\dot{M}_{\rm Edd}$, i.e.,
\begin{equation}
 \dot{M}_{\rm BH} = {\rm min}(|\dot{M}_2|, \dot{M}_{\rm Edd}),
\end{equation}
and $\dot{M}_{\rm Edd}$ is given by \citep{Pod2003},
\begin{equation}
\dot{M}_{\rm Edd}=2.6\times 10^{-7} ~M_\sun\,{\rm yr}^{-1}
\left(\frac{M_{\rm BH}}{10{M_\sun}}\right)\left(\frac{\eta}{0.1}\right)^{-1}\left(\frac{1+X}{1.7}\right)^{-1},
\end{equation}
where $X$ is the hydrogen mass fraction of the accreted material, and $\eta$
is  the efficiency of energy conversion. The excess matter is assumed to be
ejected from the binary in the form of isotropic wind.

We divide the binary evolution into two phases: the pre-CB disk phase and
the CB  disk phase. The pre-CB disk phase starts from the birth of a BH  in
a binary with a low-mass zero-age main-sequence star, and terminates
at the observed $P_{\rm orb}$, where we assume a CB disk is formed.  We try
to match the observed $P_{\rm orb}$, $M_2$ and the effective temperature
$T_{\rm eff}$ of the donor by varying the initial parameters. In the
calculation we adopt the standard AML processes including MB, GR, and mass
loss via isotropic wind. The total AML rate in the pre-CB disk phase is
given by
\begin{equation}
\dot{J}_{\rm orb}=\dot{J}_{\rm GR}+\dot{J}_{\rm MB}+\dot{J}_{\rm ML},
\end{equation}
where $J_{\rm orb}$ is the orbital AM of the binary
\begin{equation}
J_{\rm orb}=M_2M_{\rm BH}\left(\frac{Ga}{M_2+M_{\rm BH}}\right)^{1/2},
\label{jorb}
\end{equation}
and $\dot{J}_{\rm orb}$ is its time derivative.  On the right-hand-side of Eq.~(6),
$\dot{J}_{\rm GR}$ is the AML rate caused by GR \citep{Lan1975}
\begin{equation}
\dot{J}_{\rm GR}=-\frac{32}{5c^5}\left(\frac{2\pi G}{P_{\rm orb}}\right)^{7/3}
\frac{(M_2M_{\rm BH})^2}{(M_{\rm BH}+M_2)^{2/3}},
\end{equation}
where $c$ is the speed of light and $G$ is the gravitational constant. The
$\dot{J}_{\rm MB}$ term  is related to MB and described by the following
formula \citep{Ver1981,Rap1983},
\begin{equation}
\dot{J}_{\rm MB}=-3.8\times 10^{-30}M_2 R_2^4\omega_{\rm orb}^3{\rm ~dyn~cm},
\end{equation}
where  $\omega_{\rm orb}$ is the angular velocity of the binary. Finally
$\dot{J}_{\rm ML}$  represents the AML rate by isotropic wind from the BH,
\begin{equation}
\dot{J}_{\rm ML}=-(|\dot{M}_2|- \dot{M}_{\rm BH})a_{\rm BH}^2\omega_{\rm orb},
\end{equation}
where $a_{\rm BH}$ is the distance between the BH and the center of mass of the binary.

We then use the results of the pre-CB disk evolution as the input for the
calculation of the CB  disk evolution. The CB disk phase is related to the
evolution with the effect of the CB disk taken into account.  The total AML
rate in this phase is given by
\begin{equation}
\dot{J}_{\rm orb}=\dot{J}_{\rm GR}+\dot{J}_{\rm MB}+\dot{J}_{\rm ML}+\dot{J}_{\rm CB},
\end{equation}
where the $\dot{J}_{\rm CB}$ term results from the resonant interaction
between the CB disk and the  binary, and is evaluated by \citep{Lub1996}
\begin{equation}
\frac{\dot{J}_{\rm CB}}{J_{\rm orb}}=-\frac{l}{m}\frac{M_{\rm CB}}{\mu}\alpha\left(\frac{H}{R}\right)^2\frac{a}{R}\frac{2\pi}{P_{\rm orb}},
\end{equation}
where $l$ and $m$ are integers describing the binary potential, $M_{\rm CB}$
is the mass of  CB disk, $\mu$ is the reduced mass of the binary
\begin{equation}
\mu=\frac{M_2M_{\rm BH}}{M_2+M_{\rm BH}},
\end{equation}
$\alpha$ is the viscosity parameter of the disk \citep{Sha1973}, and $R$ is the half-AM
radius of the CB disk $R=\sqrt{r_{\rm in}r_{\rm out}}$ ($r_{\rm in}$: the
inner radius of the CB disk; $r_{\rm out}$: the outer radius of the CB
disk).

Our input parameters are described as follows. Assuming that the
non-axisymmetric potential  perturbations are small, we take $l=m=1$.
We adopt typical values for the viscosity in the CB disk, i.e.,
$\alpha=0.01-0.1$. \citet{Dul2001} constructed a model for irradiated dusty
CB disk. According to their study, the thickness near the inner edge of the
disk is $H/R=0.1-0.25$, so we set $(H/R)^2=1/30$. \citet{Mir2015}
investigated the tidal truncation between the CB disk and the inner binary.
Assuming the balance between resonant torque and viscous torque as the
truncation criterion, they found that the inner radius of the CB disk is
$1.5a<r_{\rm in}<3.5a$. We accordingly assume the CB disk to extend from
$r_{\rm in}=2.5a$. The outer radius of the CB disk $r_{\rm
out}$ is assumed to be $r_{\rm out}=(10-100)a$.

Recently, \citet{Jiang2017} investigated the feasibility of a CB disk in
understanding the orbital  evolution of the ultra-compact X-ray binary 4U
1820-303, which harbors a neutron star. They assume that the CB disk
originates from the material ejected by the superburst from the neutron
star. In BHLMXBs, the mass ejection feeding the CB disk may be generated by
the outburst triggered by the instability of accretion disk, or the disk
wind driven by the magnetic pressure \citep[see][for a review]{Yuan2014}.
Meanwhile, it is still under debate whether the observed orbital decay is a
short-term  or a long-term phenomenon. Thus, we take into account two cases,
in which the CB disk is fed either by one specific outburst or by a series
of outbursts. For the evolution of a debris disk, one can solve the
disk structure equations to obtain its self-similar solutions
\citep{Pringle1974,Can1990}. The total mass of the disk evolves as
\begin{equation}
M_{\rm CB}=M_{\rm CB,0}(1+\frac{t}{t_{\rm vis,0}})^{-n},
\end{equation}
where $M_{\rm CB,0}$ is the initial mass of the disk, $t_{\rm vis,0}$ is the viscous timescale at the formation of the
disk, and $n$ is a constant determined by opacity laws. Let $r_0$ be the initial radius of the disk, we can estimate the viscous timescale to be
\begin{equation}
t_{\rm vis,0}\sim \frac{r_0}{v_{\rm r}}\sim 1{\rm ~yr}\left(\frac{\alpha}{0.1}\right)^{-1}
\left(\frac{r_0}{10 a}\right)^{3/2},
\end{equation}
where $v_{\rm r}$ is the radial velocity of the disk material (here the orbital period is taken to be 0.5 d). Considering the possible range of $\alpha$ and $r_0$, w take $t_{\rm vis,0}=$1, 10, and 30~yr in our calculations.
For bound-free dominating opacities, one can obtain $n=1/4$ \citep{Can1990}. However, \citet{Ert2009} pointed out that the disk may dissipate more quickly after the the disk becomes sufficiently cool and passive. In this case there is no analytic solution the evolution of the disk mass. We tentatively adopt $n=3/4$ according to their numerically calculated results.

\section{Results}
\subsection{Pre-CB disk evolution}
In this subsection, we present the results in the pre-CB disk phase by
calculating a  number of evolutions with different initial values of $M_2$
and $P_{\rm orb}$. Several authors have theoretically investigated the
properties of the donors in BHLMXBs
\citep[e.g.,][]{Pod2003,Iva2006,Jus2006,Yun2008,Chen2015,Fro2015,Wang2016}.
In Table \ref{tab:data}, We list the observed parameters of the three
BHLMXBs. It is noted that \citet{Wu2015} obtained the effective temperatures $T_{\rm eff}$ of the donor stars by using the TDR-value method
\citep{Ton1979} to fit their spectral types,
\citet{Gon2004,Gon2006} studied the emission lines of A0620-00 and XTE
J1118-480 and obtained the effective temperatures by Monte-Carlo simulation,
and \citet{Fro2015} used an empirical relation to convert the spectral type
into the effective temperature. In our study, we adopt the values of $T_{\rm
eff}$ reported by \citet{Fro2015}, and ignore the possible X-ray irradiation
from the BH when comparing theory with observation. The reasons are that the
effect of irradiation on the donor star is still highly uncertain and all
of the three binaries are transients with very low X-ray luminosities. Table
\ref{tab:fit} lists the typical parameters selected from our calculations
for the three LMXBs after the CE phase, which can lead to the current
evolutionary stage.

\begin{figure}[!ht]
\plotone{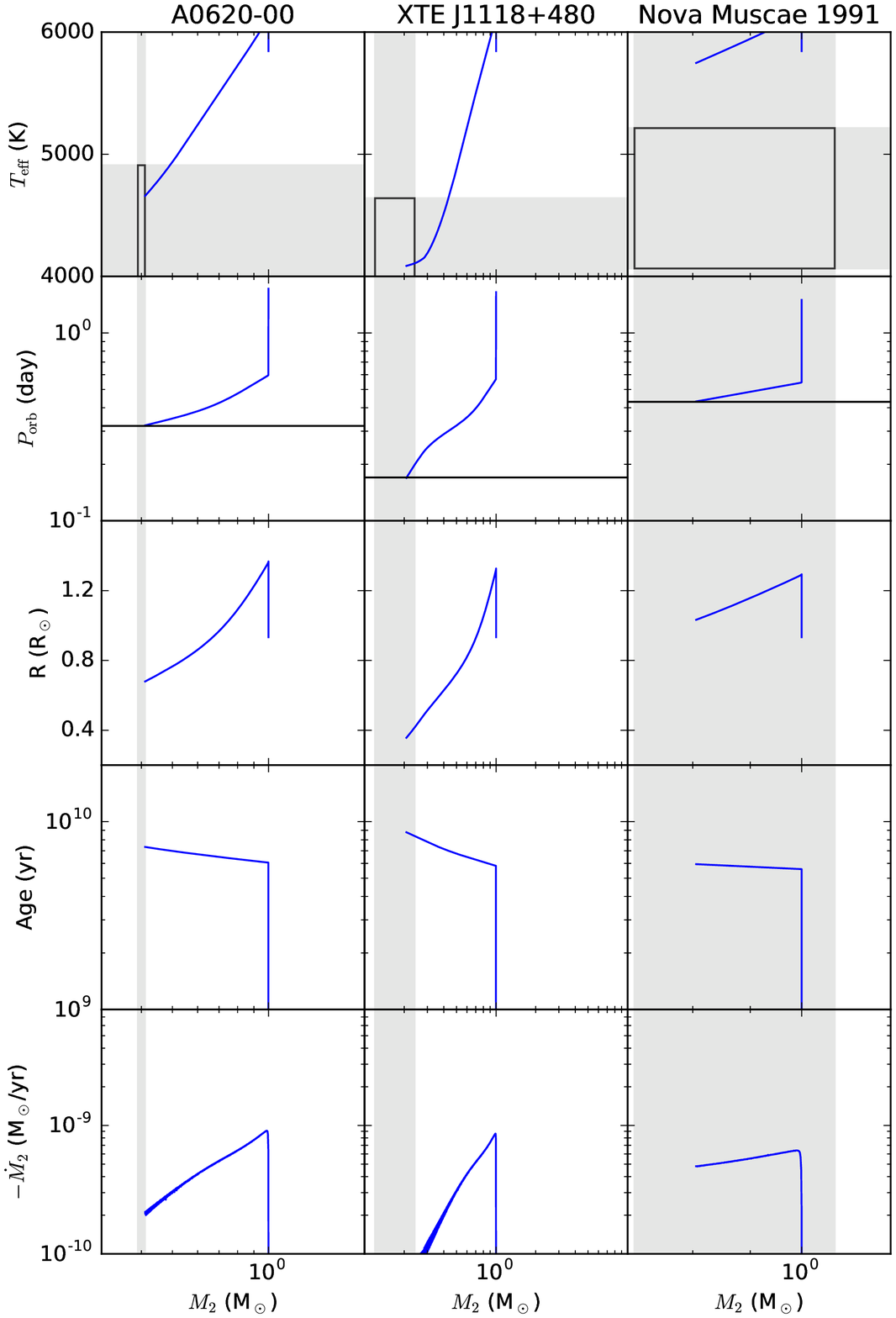}
\caption{The pre-CB disk evolution for the three BHLMXBs.
The blue solid lines represent the evolutionary tracks, and the shaded areas represent the observational constraints.
\label{f1}}
\end{figure}

Example evolutionary tracks during the pre-CB disk phase are presented in
Fig.~\ref{f1}.  The left, middle, and right rows correspond to the
evolutions of A0620-00, XTE J1118-480, and Nova Muscae 1991, respectively.
The blue solid line in each panel depicts $T_{\rm eff}$, $P_{\rm orb}$,
$R_2$, Age and $\dot{M}_2$ as a function of $M_2$.  The shaded areas, the
black rectangles, and the black horizontal lines denote the observational
constraints on $M_2$, $T_{\rm eff}$ and $P_{\rm orb}$, respectively. It can
been seen that our model can match the observations of A0620-00 and XTE
J1118+480 fairly well, but in the case of Nova Muscae 1991,  the modeled
$T_{\rm eff}$ is always slightly higher than observed.  The binaries all
show long-term orbital decay driven mainly by MB, and the mass transfer
rates are relative low ($<10^{-9}\,M_\sun$\,yr$^{-1}$).

\subsection{CB disk evolution}
\subsubsection{CB disk fed by single outburst}
\begin{figure}[!ht]
\plotone{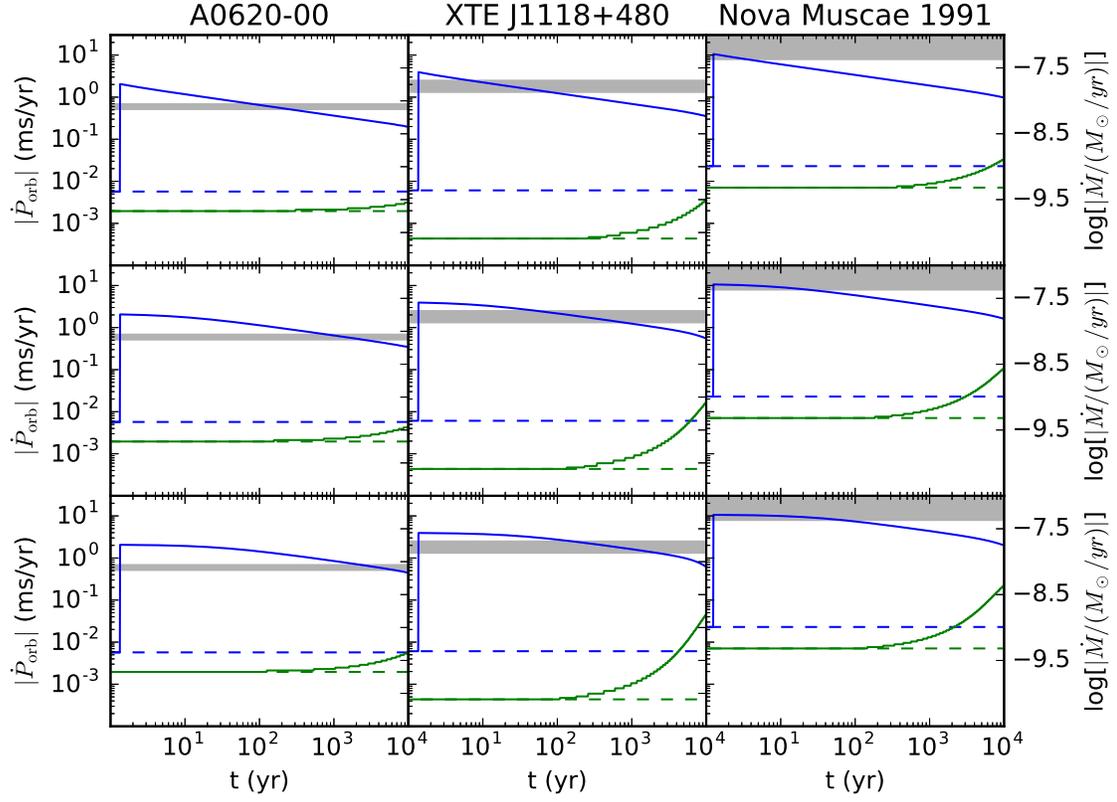}
\caption{The evolution for the three BHLMXBs with a CB disk fed by single outburst with $n=1/4$.
The input parameters $(\alpha,~r_0,~t_{\rm vis,0},~M_{\rm CB,0})$ are listed in Table~\ref{tab:f2}.
The blue and green solid lines represent the evolution of $|\dot{P}_{\rm orb}|$ and $|\dot{M}_2|$, respectively.
The dashed lines show the evolution in the absence of the CB disk for comparison.
The shaded areas represent the observed range of $|\dot{P}_{\rm orb}|$.
\label{f2}
}
\end{figure}

\begin{figure}[!ht]
\plotone{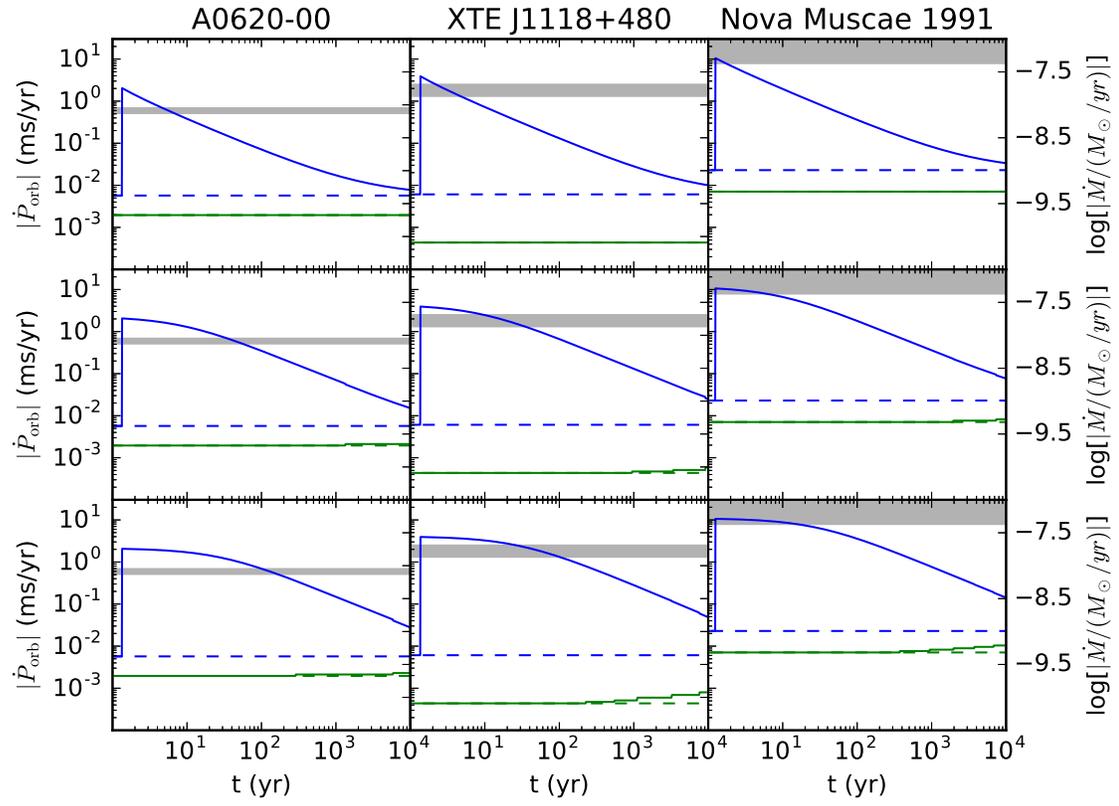}
\caption{Same as  Fig.~\ref{f2} but for $n=3/4$.
\label{f3}}
\end{figure}

In this subsection, we present the results of the evolution with the
assumption that the  CB disk is formed by mass ejection in single
outburst. Firstly we show the evolution of $|\dot{P}_{\rm orb}|$ and
log$(|\dot{M}_2|)$ during a period of $10^4$ yr in Figs.~\ref{f2} and \ref{f3}
with $n=1/4$ and $3/4$, respectively. The input parameters $(\alpha,~r_0,~t_{\rm vis,0},~M_{\rm CB,0})$ for each panel are listed in
Table~\ref{tab:f2}. The blue solid lines represent the evolution of
$|\dot{P}_{\rm orb}|$ with $\dot{P}_{\rm orb} < 0$, and the green solid
lines the evolution of $|\dot{M}_2|$, respectively. For comparison, we also
plot the evolution of $|\dot{P}_{\rm orb}|$ and $|\dot{M}_2|$ in the absence of a CB disk with the blue and green dashed lines, respectively. The shaded
areas denote the observed range of $\dot{P}_{\rm orb}$.

The general feature in Figs.~\ref{f2} and \ref{f3} is that, when the system
enters the CB disk phase,  the value of $-\dot{P}_{\rm orb}$ increases
immediately, and then decreases roughly on a viscous timescale.
Theoretically it is possible to explain the rapid orbital shrinkage of the
BHLMXBs with reasonable values of $M_{\rm CB,0}$ and $t_{\rm vis,0}$. It is  known that a rapid
orbital decay can accelerate mass loss from the donor, consequently enhance
the X-ray luminosity \citep{Jus2006,Chen2015}. However, observationally the
three BHLMXBs are all faint sources. Figs.~\ref{f2} and \ref{f3} show that
there is a time delay between the increase in $|\dot{P}_{\rm orb}|$ and
$|\dot{M}_2|$, which means that it takes the donor a relaxation time to
adjust its structure in response to the increase in $|\dot{P}_{\rm orb}|$. In the case of $n=3/4$ the change of $|\dot{M}_2|$ is much smaller than in the case of $n=1/4$ because of more rapid disk dissipation.
Thus, it is possible that the formation of CB disk can lead to a rapid
orbital decay without changing $|\dot{M}_2|$ significantly.

\begin{figure}[!ht]
\plotone{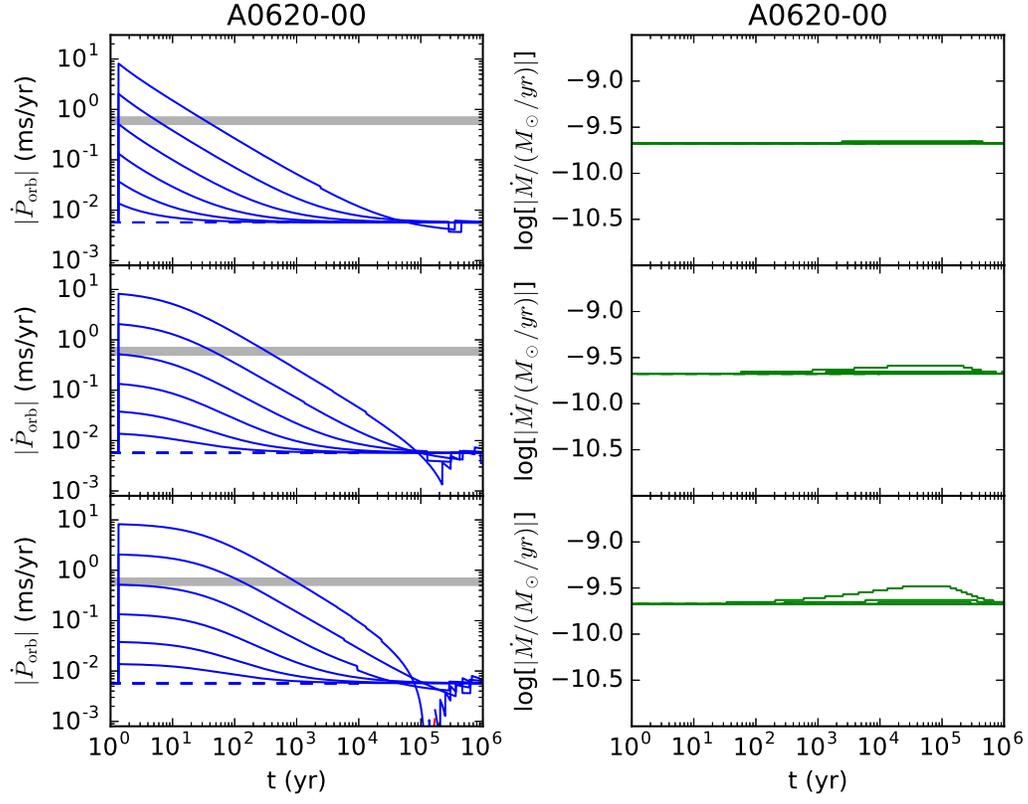}
\caption{
The evolution with a CB disk for A0620-00.
In each panel the solid curves from top to bottom are obtained with $\zeta=(4,~1,~4^{-1},~4^{-2},~4^{-3},~4^{-4})$, respectively.
Other symbols have the same meanings as in Fig.~2.
\label{f4}}
\end{figure}

\begin{figure}[!ht]
\plotone{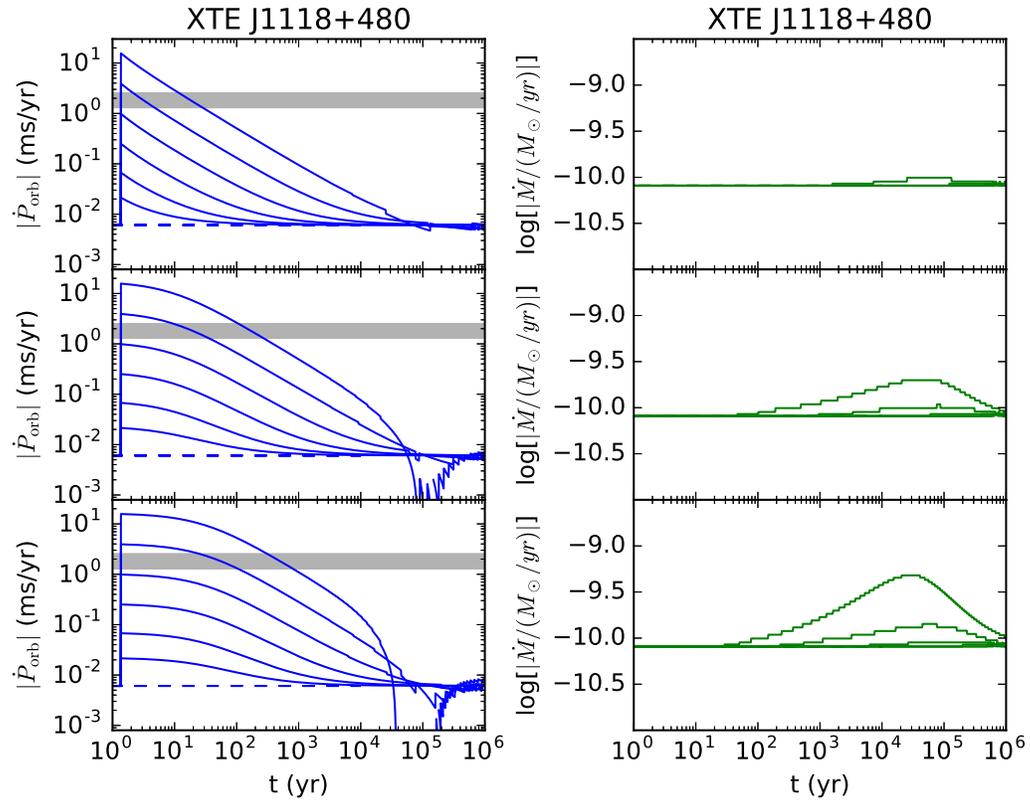}
\caption{
Same as Fig.~\ref{f4} but for  XTE J1118+480.
\label{f5}}
\end{figure}

\begin{figure}[!ht]
\plotone{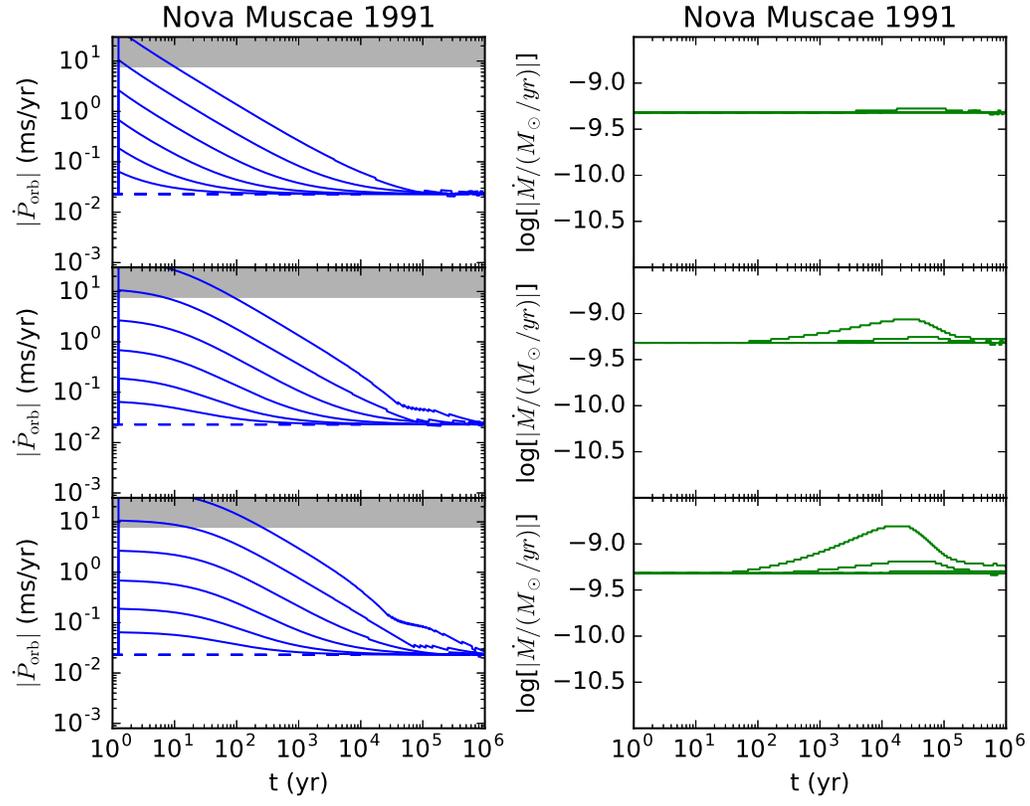}
\caption{
Same as Fig.~\ref{f4} but for Nova Muscae 1991.
\label{f6}}
\end{figure}

Then, we study the effect of $M_{\rm CB}$ on the CB disk evolution within a
relatively long  time of $10^{6}$ yr. Here we take $n=3/4$.  The
results are shown in Figs.~\ref{f4}, \ref{f5}, and \ref{f6} for A0620-00,
XTE J1118+480, and Nova Muscae 1991, respectively. In each figure, the left
and right columns depict the evolution of $|\dot{P}_{\rm orb}|$ (blue solid
and dashed lines for $|\dot{P}_{\rm orb}|$ with and without a CB disk respectively) and $|\dot{M}|$. For each BHLMXB, we
adopt the same parameters in Table~\ref{tab:f2} but adding a factor $\zeta$  to $M_{\rm CB,0}$ with $\zeta=(4,~1,~4^{-1},~4^{-2},~4^{-3},~4^{-4})$.
corresponding to the six evolutionary tracks in each panel from top to
bottom, respectively.

In the case of A0620-00, the CB disk can lead to orbital decay for a time of a few $10^4$ yr. The orbital decay with $\zeta=(4,~1,~4^{-1})$ is
fast enough to  match the observed value. The more massive the CB disk and the longer the initial viscous time, the
longer time for the orbital shrinkage, consequently the higher mass loss rate.
In the case of $t_{\rm vis,0}=30 $ yr and $\zeta=4$, the mass transfer can even cause the orbital period to expand for some time, which is depicted with the red line in the lower left panel. After that the orbital period decreases again due to AML
by MB and GR.

The general picture of the evolution of XTE J1118+480 is similar to that of
A0620-00. The donor  of XTE J1118+480 is older and less massive, and MB in
the donor has already been turned off because of the absence of the
radiative core. Thus, the mass loss process of the donor is more sensitive
to the change in $\dot{P}_{\rm orb}$.  Before the formation of the CB disk,
$|\dot{M}_2|$ is very low ($< 10^{-10}\,M_\sun$\,yr$^{-1}$). When entering
the CB disk phase, the rapid orbit shrinkage causes a higher enhancement of
$|\dot{M}_2|$ than in A0620-00. Compared with the other two sources, the
donor of Nova Muscae 1991 is younger and more massive. Thus, before the
formation of the CB disk, $|\dot{M}_2|$ is relatively high. The donor
responds more quickly to the change in $\dot{P}_{\rm orb}$, leading to
shorter duration of the $|\dot{M}_2|$-enhancement.

\subsubsection{CB disk fed by successive outbursts}
\begin{figure}[!ht]
\plotone{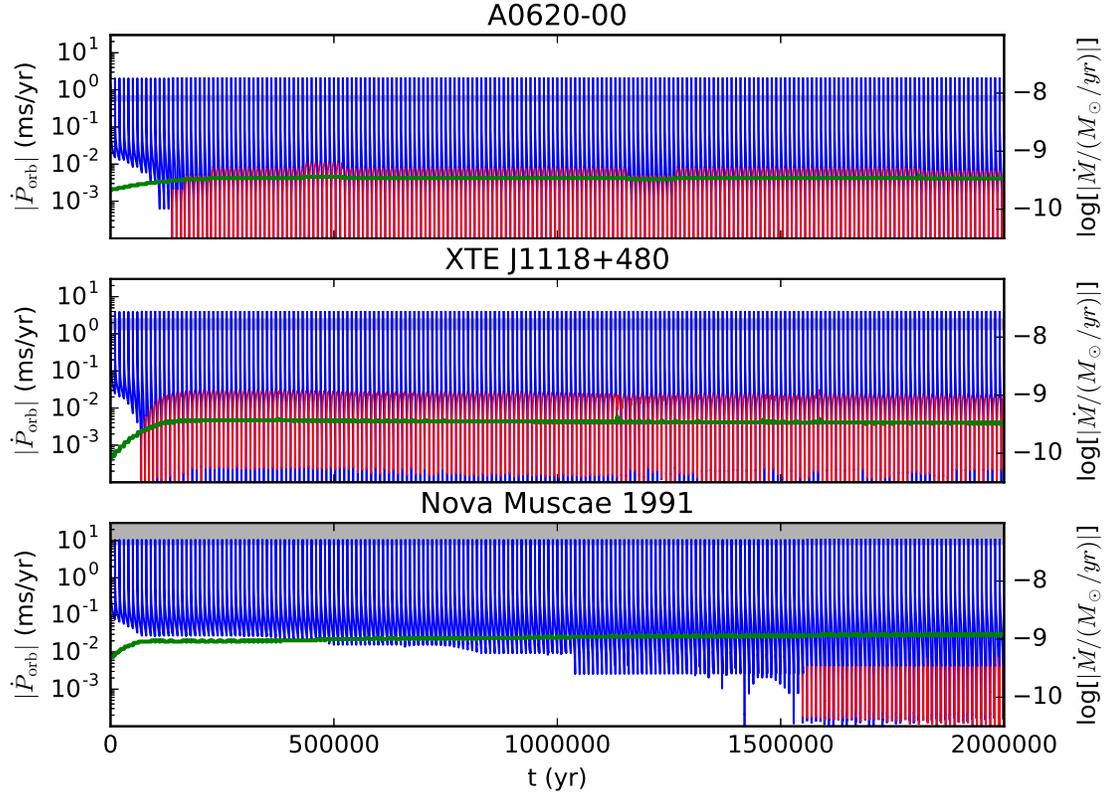}
\caption{
The evolution with a CB disk fed by repeating outbursts.
Other symbols have the same meanings as in Fig.~2.
\label{f7}}
\end{figure}

\begin{figure}[!ht]
\plottwo{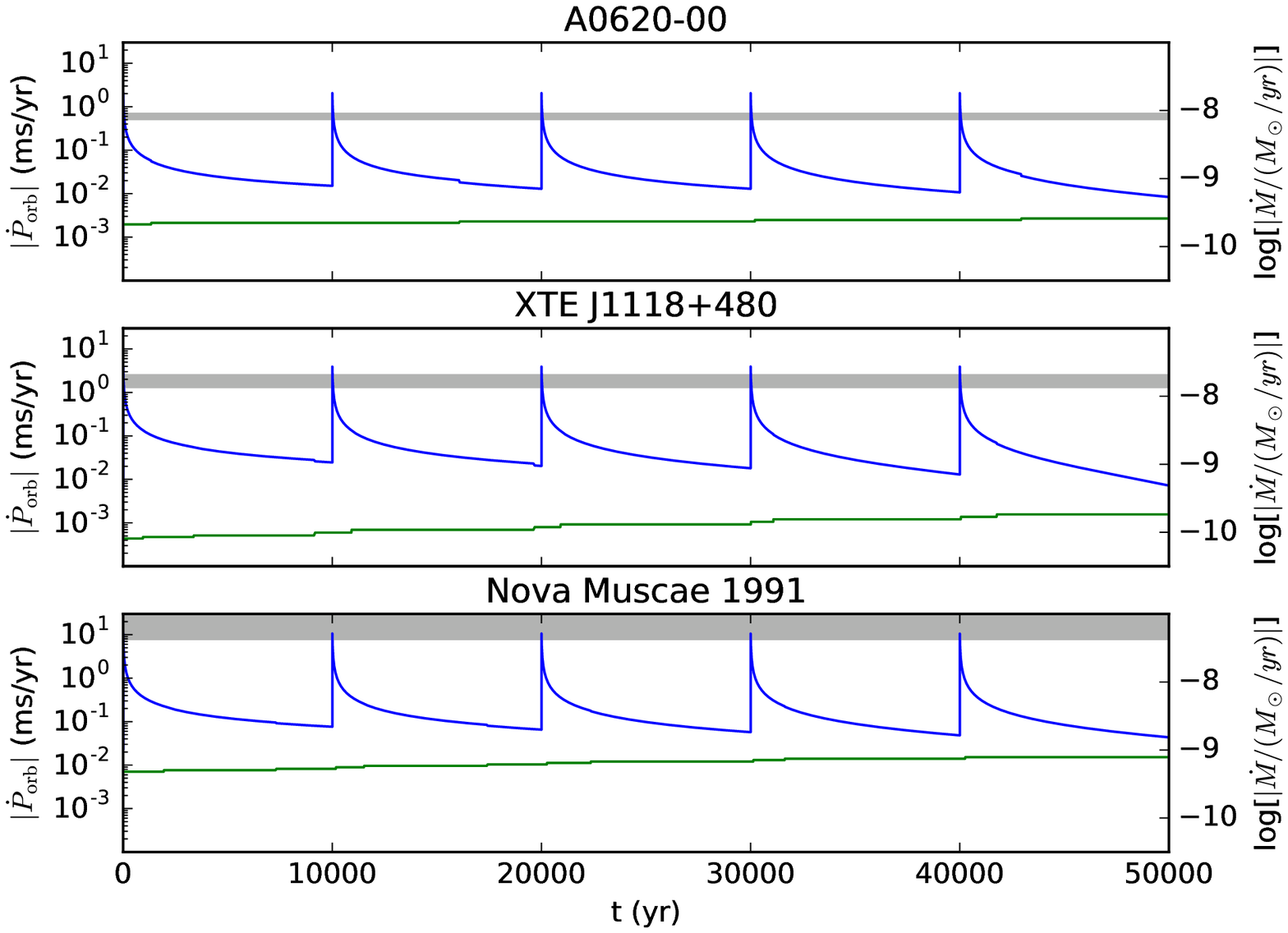}{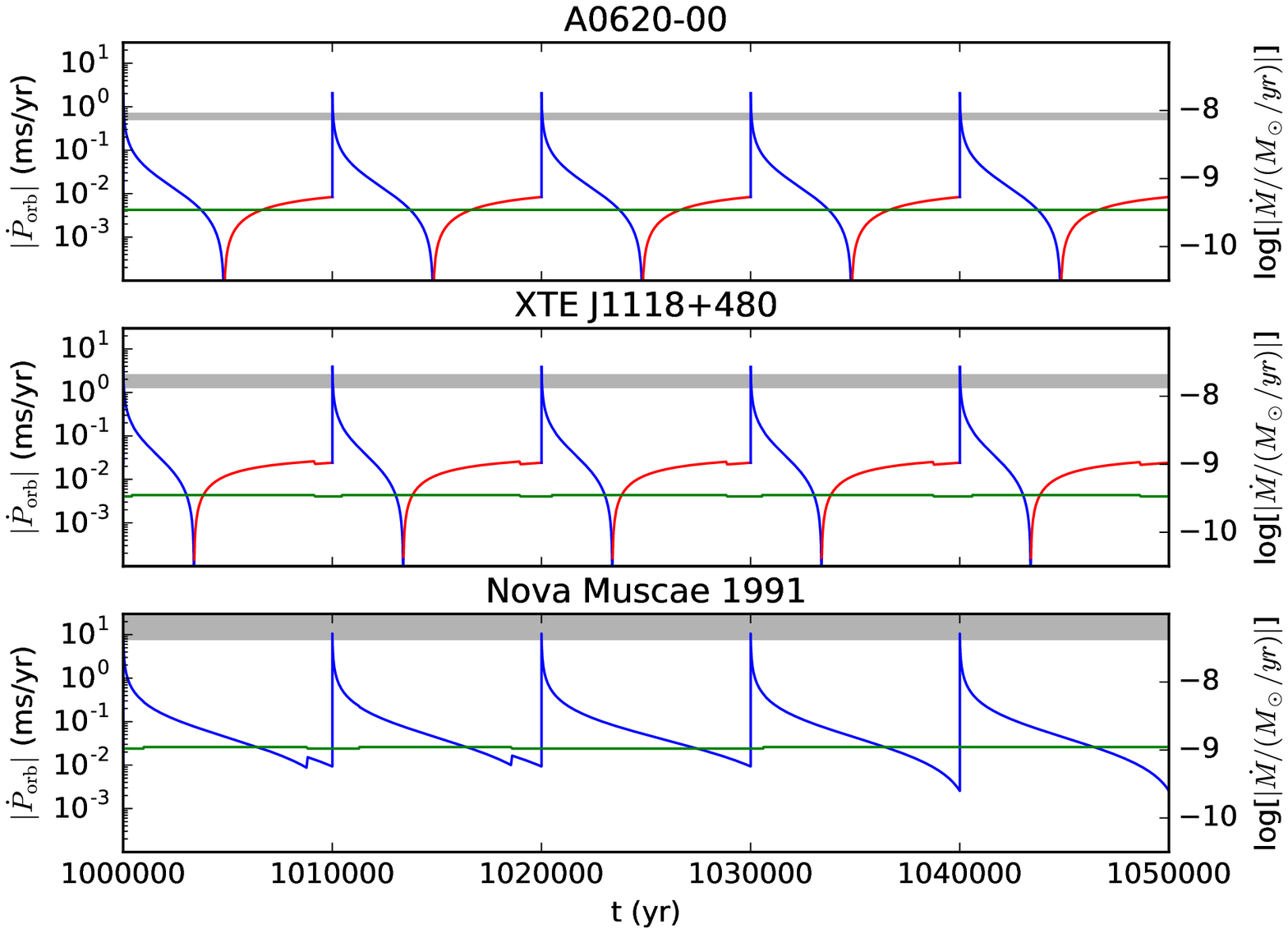}
\caption{
Zoom in of Fig. \ref{f7} during the time 0 - 50000 yr (left) and
$10^6$ yr - ($10^6+50000$) yr (right).
\label{f8}
}
\end{figure}

In this subsection we consider the alternative case that the CB disk is fed
by a series of successive mass ejection processes from the binary. We assume
that the lost material during the outbursts with a duty cycle of $1/200$, which is typical in BHLMXBs, is repeatedly added to the CB disk, which then dissipates
on a viscous timescale. As an illustration, we present the evolution of $|\dot{P}_{\rm orb}|$ and $|\dot{M}_2|$ within $2\times 10^6$ yr in
Fig.~\ref{f7}. Here we adopt the input parameters in the
middle row in Table~\ref{tab:f2} with $n=3/4$.  The upper, middle, and lower panels depict the results for
A0620-00, XTE J1118+480, and Nova Muscae 1991, respectively. The blue and
red solid lines represent the evolution of $|\dot{P}_{\rm orb}|$ with
$\dot{P}_{\rm orb} < 0$ and $> 0$ respectively, and the green line
represents the evolution of $|\dot{M}_2|$. The shaded areas denote the
observed range of $\dot{P}_{\rm orb}$. A zoom-in of the features in
Fig.~\ref{f7} is shown in Fig.~\ref{f8}, where the left and right panels
demonstrate the evolution in  $0 - 50000$ yr and in $10^6$  -
($10^6+50000$) yr, respectively.

As we demonstrate in last subsection, in the case of single outburst-induced CB disk, there is a time
delay between the  increases in $|\dot{P}_{\rm orb}|$ and in $|\dot{M}_{2}|$, and
after the dissipation of the CB disk the donor takes a relatively long
relaxation time ($>10^4$ yr) to return to the previous evolutionary path.
For successive outbursts
the donor does not have enough time to adjust its structure in responding to
the dissipation of the CB disk before the next cycle of mass feeding.
Consequently, $|\dot{M}_2|$  increases within the initial several cycles,
during which the orbital period keeps decreasing. At some point, $\dot{M}_2$
is high enough to induce orbital expansion after the dissipation of the CB
disk in each cycle,  which limits the increasing trend of $|\dot{M}_2|$. The
orbital shrinkage and expansion take place in turn, and $|\dot{M}_2|$
reaches a relatively stable value.  For A0620-00, XTE J1118+480, and Nova
Muscae 1991, it is $<5\times 10^{-10}\,M_{\sun}$\,yr$^{-1}$,
$<5\times 10^{-10}\,M_{\sun}$\,yr$^{-1}$, and $<1.5\times
10^{-9}\,M_{\sun}$\,yr$^{-1}$, respectively. From Figs.~\ref{f7} and
\ref{f8}, the time percentages for the orbital decay rates larger than
observed ones are $0.50\%$, $0.31\%$ and $0.046\%$  for A0620-00, XTE
J1118+480, and Nova Muscae 1991, respectively. Obviously these values can be
increased considerably if taking larger $t_{\rm vis,0}$ and/or smaller $n$. Therefore, a
CB disk fed by successive outburst may also explain the rapid orbital
shrinkage while keeping $|\dot{M}_2|$ relatively low.

\begin{figure}[!ht]
\plotone{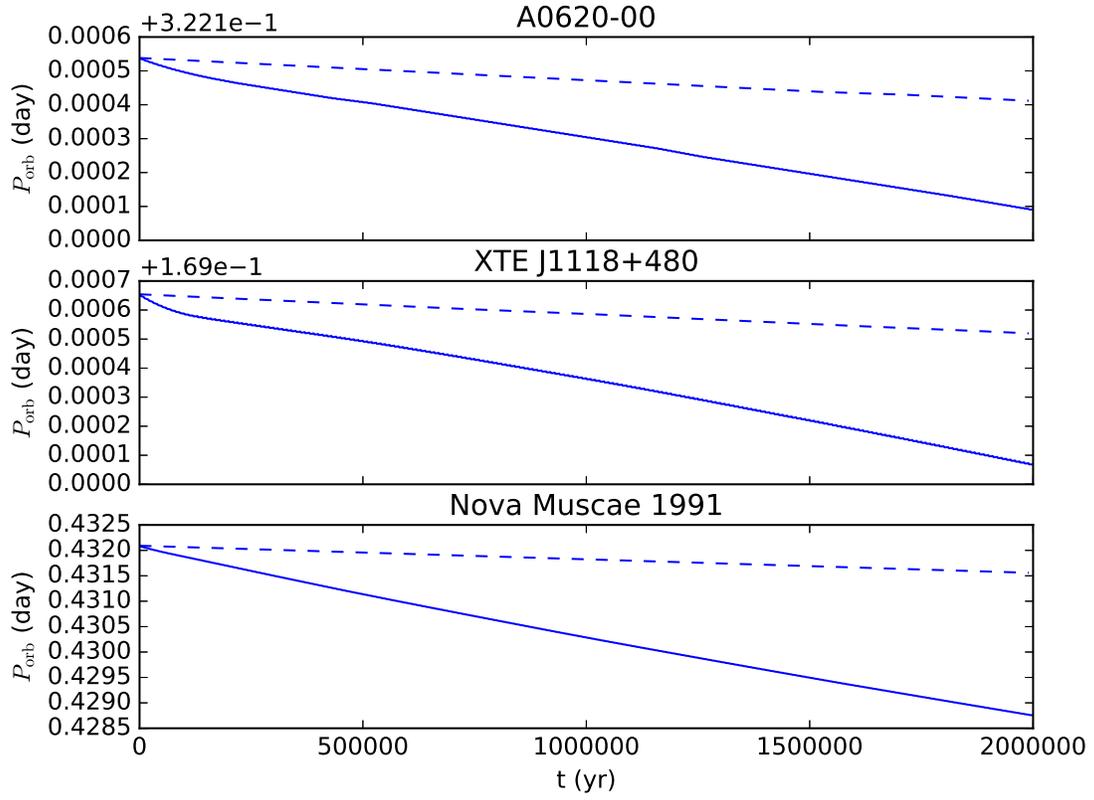}
\caption{
Comparison of the long-term orbital evolution with and without a CB disk, depicted with the solid and dashed lines, respectively.
\label{f9}}
\end{figure}

We also plot the evolution of $P_{\rm orb}$ in Fig.~\ref{f9}, in which the
upper, middle,  and lower panels correspond to A0620-00, XTE J1118+480, and
Nova Muscae 1991, respectively. The blue solid and dashed lines represent
the evolution with and without a CB disk. This figure shows that, although
both orbital shrinkage and expansion take place in the CB disk case, the
former dominates the long-term evolution, and results in a orbital decay
faster than driven only by MB and GR.

\section{Discussion}

The rapid orbital shrinkage in the three BHLMXBs remains to be a puzzle in the traditional framework of LMXB evolution.  Besides the dynamical interaction with a third-body,
one of the most natural explanations is efficient AML associated with mass
loss. However, this scenario usually induces rapid mass transfer at a rate
higher than observed. Meanwhile, infrared observations point to the possible
existence of a CB disk around the binaries but strongly constrain its mass
\citep{Mun2006}.

In this work we take into account the effect of the resonant interaction
between the  CB disk and the inner binary, assuming that the CB disk
originates from outburst(s) triggered by some instability in the BH
accretion disk. Because the origin and evolution of the CB disk is still
uncertain, we consider that the disk is fed by either single outburst or
successive outbursts, and examine how the interaction changes with initial
mass and lifetime  of the disk. Our results show that the resonant
interaction between the CB disk and  the binary may explain the observed
$\dot{P}_{\rm orb}$ with the disk mass comparable with observations and no
significant change in $\dot{M}_2$.

In this work we assume that the CB disk originates from mass ejection during outburst(s) which may be caused by thermal instability in the accretion disk around the BH when the accretion rate is lower than a critical value \citep{Las2001}. Under this condition the successive outburst model may be more viable than the single outburst model since all transient BHLMXBs are believed to be recurrent. However, it is difficult to definitely determine when the disk starts to be unstable and how much mass ejected during a specific outburst goes into the CB disk, so we arbitrarily assume that the CB disk is formed at current time. With this approach we can focus on the effect of the CB disk on the orbital evolution with the binary parameters compatible with observations.
If the CB disk was formed at an earlier time, the long-term (CB disk-assisted) orbital decay would proceed at a higher rate than only with MB and GR taken into account, then a smaller initial donor mass would be required.

For the disk-binary interaction, dynamical friction was also
proposed in the literature  \citep[e.g.,][]{Sye1995, Iva1999}. Under this
circumstance, the secondary is assumed to be embedded and open a gap in an accretion disk surrounding the primary. The width of the gap is assumed to remain considerably smaller than the orbital radius of the secondary to guarantee dynamical interaction. It was found that the timescale of orbital shrinkage is much longer than the local viscous
timescale of the disk if the disk mass is significantly smaller than the secondary's mass. The condition is different for the CB disk case. Here the CB disk, with a central cavity, is coupled with the inner binary via long-range resonant interaction \citep{Gol1979,Lub1997}.
\citet{Raf2016} recently investigated the angular momentum exchange between a post-main-sequence binary and a viscously evolving CB disk.
It can be shown that the torque acting on the binary derived by \citet{Raf2016} is consistent with Eq.~(12) in our work. Therefore, while the mass
of CB disk is significantly lower than that of the secondary, it is capable of efficiently
extracting angular momentum via resonant interaction.

We note that orbital changes have also been observed in neutron star
LMXBs.  For example, SAX J1808.4-3658, 2A 1822-371, and SAX J1748.9-2021
show orbital expansion \citep{Har2008,Bur2010,San2016}, while X1658-298 and
AX J1745.6-2901 \citep{Wac2000,Pon2016} show orbital shrinkage. Since these
LMXBs are all short-period binaries that share some similarities with the
three BH binaries, the implication is that more than one kinds of mechanisms
might be at work in LMXBs. To account for the observed period change, the
Applegate mechanism requires a $\sim 1$ KG surface magnetic field for the
donor star, and the CB disk model predicts that the eccentricity could
evolve with the orbital change \citep{Der2013,Ant2014,Raf2016}. More detailed observations are expected to
provide stringent constraint on the binary evolution and discriminate
various kinds of models.

\acknowledgments We are grateful to the referee for helpful comments. This work was supported by he National Key Research and
Development Program of China  (2016YFA0400803), and the Natural Science
Foundation of China under grant numbers 11773015, 11133001 and 11333004.

\newpage

\newpage

\newpage

\newpage

\newpage

\newpage

\newpage
\begin{deluxetable}{l cccccc}[b!]
\tablecaption{Observed parameters of the BHLMXBs \label{tab:data}}
\tablecolumns{6}
\tablenum{1}
\tablewidth{0pt}
\tablehead{
\colhead{Parameter\tablenotemark{a}} &\colhead{Nova Muscae 1991} &\colhead{Ref.\tablenotemark{b}} &
\colhead{XTE J1118+480} & \colhead{Ref.\tablenotemark{b}} &
\colhead{A0620-00} & \colhead{Ref.\tablenotemark{b}}
}
\startdata
$M_{\rm BH}$ (M$_\sun$)& $11.0^{+2.1}_{-1.4}$&[1]&$7.46^{+0.34}_{-0.69}$&[2]&$6.61^{+0.23}_{-0.17}$&[2]\\
$M_2$ (M$_\sun$)&  $0.89\pm 0.18$&[1]& $0.18\pm0.06$&[2]&$0.40\pm0.01$&[2]\\
$q=M_2/M_{\rm BH}$ & $0.079\pm0.007$& [3]& $0.024\pm0.009$&[2]&$0.06\pm0.004$&[4]\\
$a$ (R$_\sun$)&$5.49\pm0.32$&[1]&$2.54\pm0.06$&[2]&$3.79\pm0.04$&[2,5]\\
$R_2$ (R$_\sun$)&$1.06\pm0.07$&[1]&$0.34\pm0.05$&[2]&$0.67\pm0.02$&[2,5]\\
$P_{\rm orb}$ (day)&0.432605(1)&[6]&0.16993404(5)&[2]&0.32301415(7)&[2]\\
$\dot{P}_{\rm orb}$ (ms yr$^{-1}$)&$-20.7\pm12.7$&[6]&$-1.90\pm0.57$&[2]&$-0.60\pm0.08$&[2]\\
$T_{\rm eff} (K)$&$4400\pm100$&[3]&$4700\pm 100$&[8]&$4900\pm100$&[7]\\
 &4065-5214&[9]&3405-4640&[9]&3800-4910&[9]\\
\enddata
\tablenotetext{a}{Meanings of the parameters: $M_{\rm BH}$ - the BH mass, $M_2$ - the donor mass, $q$ - the mass ratio,
$a$ - the semi-major axis of the orbit, $R_2$ - donor's radius, $P_{\rm orb}$ - the orbital period,
$\dot{P}_{\rm orb}$ - the changing rate of the orbital period, $T_{\rm eff}$ - the effective temperature of the donor.}
\tablenotetext{b}{References: [1]\citet{Wu2016}; [2]\citet{Gon2014}; [3]\citet{Wu2015};
[4]\citet{Nei2008}; [5]\citet{Gon2011}; [6]\citet{Gon2017};
[7] \citet{Gon2004}; [8] \citet{Gon2006}; [9] \citet{Fro2015}
}
\end{deluxetable}

\begin{deluxetable}{l cccccc}[hb!]
\tablecaption{Selected initial parameters for the BHLMXBs \label{tab:fit}}
\tablecolumns{4}
\tablenum{2}
\tablewidth{0pt}
\tablehead{
\colhead{Parameter} &\colhead{Nova Muscae 1991}&\colhead{XTE J1118+480} & \colhead{A0620-00}
}
\startdata
$M_2$ (M$_\sun$) &        1  & 1.0& 1\\
$M_{\rm BH}$ (M$_\sun$) & 11 & 6.6 & 6\\
$P_{\rm orb} (day)$ &    1.5& 1.65& 1.72 \\
\enddata
\end{deluxetable}

\begin{deluxetable}{l cccccc}[hb!]
\tablecaption{Input parameters $(\alpha,~r_0,~t_{\rm vis,0},~M_{\rm CB,0})$ \label{tab:f2}}
\tablecolumns{4}
\tablenum{3}
\tablewidth{0pt}
\tablehead{
\colhead{} &
\colhead{A0620-00}&\colhead{XTE J1118+480} & \colhead{Nova Muscae 1991}
}
\startdata
upper row&$(0.1,~10a,~1{\rm ~yr},~4\times10^{24}{\rm ~g})$&$(0.1,~10a,~1{\rm ~yr},~4\times10^{24}{\rm g})$&
$(0.1,~10a,~1{\rm ~yr},~4\times10^{25}{\rm g})$\\
middle row&$(0.01,~10a,~10{\rm ~yr},~4\times10^{25}{\rm g})$&$(0.01,~10a,~10{\rm ~yr},~4\times10^{25}{\rm g})$&
$(0.01,~10a,~10{\rm ~yr},~4\times10^{26}{\rm g})$\\
lower row&$(0.1,~100a,~30{\rm ~yr},~12.8\times10^{24}{\rm g})$&$(0.1,~100a,~30{\rm ~yr},~12.8\times10^{24}{\rm g})$&
$(0.1,~100a,~30{\rm ~yr},~12.8\times10^{25}{\rm g})$\\
\enddata
\end{deluxetable}

\end{document}